# Surface scattering velocities in III-Nitride quantum well laser structures via the emission of hybrid phonons.

V. N. Stavrou[1,2] and G.P. Veropoulos[1]
1. *Division of Academic Studies, Hellenic Navy Petty Officers Academy, Skaramagkas, T.K. 12400, Greece*
2. *Department of Physics and Astronomy, University of Iowa, Iowa City, IA 52242, USA*

**Abstract**

We have theoretically and numerically studied nitride-based quantum well (QW) laser structures. More specifically, we have used a QW made with III-nitride where the width of the barrier region is large relative to the electron mean free path and we have calculated the electron surface capture velocities by considering an electron flux which is captured into the well region. The process is assisted by the emission of the longitudinal optical (LO) phonons as predicted by the hybrid (HB) model. The results of surface capture velocities via the emission of HB phonons are compared to the emission of the dielectric continuum (DC) phonons [Ref. 16]. Our investigation shows that the two different phonon models predict almost the same results for the non-retarded limit. Furthermore, the surface capture velocities strongly depend on the size of the structure and the heterostructure materials. Lastly, comparison to the recent experimental values shows that our model could accurately describe experimentally measured parameters of the quantum capture processes.

**PACS:** 63.20.Dj, 72.10.Di, 72.20.Jv

**Keywords:** Semiconductor Lasers, large energy gaps, GaN, hybrid phonons, surface capture velocity.

## 1. Introduction

During the last decades, semiconductor industry has focused its interest on growth QW lasers structures made, among others, with III-nitride materials [1-14]. Low dimensional structures (LDS) based on nitride semiconductor materials have been used for fabricating light emitting diodes. Due to their large energy gaps, III-nitride materials can be used for emitters which are active in the blue/green and ultraviolet region as well. In recent years, several techniques have been employed to fabricate high quality QW heterostructures [1-3,13-14] and experimental results have been reported concerning the ability of QWs for lasing [10-12].

The mechanism of electron capture by heterostructures is of fundamental importance to improve the performance of semiconductor laser. The electron capture process (the electron capture rate and capture velocity) [15-16] is mediated by the longitudinal optical phonons (LO). In theoretical studies of the capture mechanism, two different approaches have been used in order to describe the initial electron wavefunctions in the continuum energy spectrum of the QW which captured. The first approach considers that the width of the barrier region is small relative to the electron mean free path [15] and the second approach is opposite [16]. Phonons models which have successfully described the LO phonons in heterostructures among others are the DC model [17-18] and the HB model [19-21]. The DC model considers only the electromagnetic boundary conditions (BCs) and can successfully describe only the non-retarded limit. The HB model uses mechanical and electromagnetic boundary conditions which can predict both the effects in the Raman region and the non-retarded limit [19-20].

In this paper, the surface capture velocities have been estimated for the case of an electron flux which is captured in the quantum well (where the barrier width is large relative to the electron mean free bath) by the emission of HD phonons. The comparison with the DC model shows that these two different models predict almost the same capture velocities. The recently reported experimental results on capture times are comparable to our theoretical calculations. This paper is arranged as follows: In § 2, we describe the theoretical model regarding the phonon modes which is suitable for the case of the III-nitrides. We also analyse the wavefunctions of the initial/final electron states and introduce the capture velocities using the Fermi Golden Rule. The results and the conclusions are respectively given in § 3 and § 4.

## 2. Theory

In this section, we firstly describe the hybrid phonon modes and their boundary conditions (BCs) for a QW structure as is illustrated in Fig.1. The equation of motion for the ionic relative displacement field $\mathbf{u}$, including the bulk dispersion of LO and TO modes, in terms of the electric field $\mathbf{E}$ and the polarization field $\mathbf{P}$ is given by the following equations:

$$\ddot{\mathbf{u}} = -\omega_T^2 \mathbf{u} + \varepsilon_0 \varepsilon_\infty \rho^{-1} \tilde{e} \mathbf{E} - v_L^2 \nabla(\nabla \cdot \mathbf{u}) + v_T^2 \nabla \times (\nabla \times \mathbf{u}) \quad (1)$$

$$\mathbf{P} = \varepsilon_0 \varepsilon_\infty \tilde{e} \mathbf{u} + \varepsilon_0 (\varepsilon_\infty - 1) \mathbf{E} \quad (2)$$

$$\tilde{e}^2 = \frac{\rho(\omega_L^2 - \omega_T^2)}{\varepsilon_0 \varepsilon_\infty} \quad (3)$$

where $\tilde{e}$, $\varepsilon_0$, $\varepsilon_\infty$ and $\rho$ are the effective ionic charge, the permittivity of free space, the high-frequency dielectric constant and the reduced mass density for the ions respectively. In equation (1) the last two terms have been introduced to describe the effect of bulk spatial dispersion, with $v_L$ and $v_T$ as velocity parameters [19]. It is necessary to make clear at this point that **u** is the ionic relative displacement and is related to the renormalized displacement **w** by

$$\mathbf{u} = \frac{\mathbf{w}}{\sqrt{\rho}} \tag{4}$$

The electric displacement field in terms of electric field **E** and the ionic relative displacement field **u** is given by

$$\mathbf{D} = \varepsilon_0 \varepsilon_\infty (\mathbf{E} + \tilde{\mathbf{e}}\mathbf{u}) \tag{5}$$

It is worth mentioning that the electric displacement can not be defined at a given value of z [15, 19]. This property is generally referred to as nonlocality and is responsible for the mechanical boundary conditions which need to be taken into account in the HB model.

The general solution of the hybrid modes is the sum of the longitudinal $\mathbf{u}_L$, the transverse $\mathbf{u}_T$ and the interface $\mathbf{u}_I$ part. The resultant fields are called triple hybrids:

$$\mathbf{u} = \mathbf{u}_L + \mathbf{u}_T + \mathbf{u}_I \tag{6}$$

where the longitudinal, the transverse and the interface parts are respectively given by

$$\begin{aligned} \nabla \times \mathbf{u}_L &= 0 \\ \nabla \cdot \mathbf{u}_T &= 0 \\ \nabla \times \mathbf{u}_I &= 0, \; \nabla \cdot \mathbf{u}_I = 0 \end{aligned} \tag{7}$$

The TO part of the hybrid mode does not have any electric field and so it does not contribute to the Fröhlich coupling mechanism [15, 20]. Hence, for the calculations of scattering rates, the TO part can be removed.

The hybrid modes satisfy both electromagnetic BCs and mechanical BCs. Thus, the electromagnetic conditions demand the continuity of the in-plane electric field component and the continuity of the normal component of the electric displacement. The mechanical conditions demand that the normal components of the stress tensor [19-20] be continuous. This can be shown to amount to the continuity of the components of **u** which are tangential and perpendicular to any interface [21]. The Hamiltonian for a hybrid mode in terms of ionic displacement operator $\hat{\mathbf{u}}(\mathbf{r})$ of the hybridons, the crystal volume $V_c$ and the reduced mass M is given by [20]

$$\hat{H} = \frac{M}{2V_c}\left[\int \dot{\hat{u}}^2(\mathbf{r})d^3\mathbf{r} + \omega^2 \int \hat{u}^2(\mathbf{r})d^3\mathbf{r}\right] \tag{8}$$

In the case of III-nitride materials, the mechanical displacement modes from QW material may also propagate in the barrier region as well because the reststrahl bands overlap in the first Brillouin zone [21]. As a result, we can not use the approximation which has been applied for the case of GaAs/AlAs [15a].

In the second part of our investigation, we consider that an initial wave is propagating from $z = -\infty$ (Fig. 1) and that the wavefunction is normalised to a unit carrier density $j_z$:

$$j_z = \frac{i\hbar}{2m_2^*}\left[\Psi(\mathbf{r})\frac{d\Psi^*(\mathbf{r})}{dz} - \Psi^*(\mathbf{r})\frac{d\Psi(\mathbf{r})}{dz}\right] \qquad (9)$$

Using eq. (9), the initial electron wavefunction could be described by

$$\Psi_k(z) = A \begin{cases} e^{ik_2 z} + R_0 e^{-i\left[k_2(z+2L)-\theta+\frac{\pi}{2}\right]} & z \leq -L \\ P_0 e^{i[k_1(z-L)-k_2 L+\theta]} + Q_0 e^{-i[k_1(z-L)+k_2 L-\theta]} & |z| \leq L \\ S_0 e^{i[k_2(z-2L)+\theta]} & z \geq L \end{cases} \qquad (10)$$

where $|A|$ is defined as $|A|^2 = (2\pi)^2 \frac{m_2^* j_z}{\hbar k_2}$. The coefficients $R_0$, $P_0$, $Q_0$ and $S_0$ can be defined by the boundary conditions (continuity of $\psi(z)$ and $\frac{1}{m^*}\frac{d\psi}{dz}$) at the interfaces. Using the boundary conditions, the coefficients $S_0$, $R_0$, $P_0$ and $Q_0$ are respectively given by

$$S_0 = T^{\frac{1}{2}} \qquad R_0 = \frac{1}{2}T^2\left(\frac{m_2^* k_1}{m_1^* k_2} - \frac{m_1^* k_2}{m_2^* k_1}\right)\sin(2k_1 L) \qquad (11)$$

$$P_0 = \frac{1}{2}T^2\left(1 + \frac{m_1^* k_2}{m_2^* k_1}\right) \qquad Q_0 = \frac{1}{2}T^2\left(1 - \frac{m_1^* k_2}{m_2^* k_1}\right) \qquad (12)$$

Only a part of the wave is transmitted and it is described by the transmission coefficient

$$T = |S_0|^2 = \frac{1}{\cos^2(2k_1 L) + \frac{1}{4}\left(\frac{m_2^* k_1}{m_1^* k_2} + \frac{m_1^* k_2}{m_2^* k_1}\right)^2 \sin^2(2k_1 L)} \qquad (13)$$

The corresponding reflection coefficient R, is given by

$$R = |R_0|^2 = \frac{1}{4}|s_0|^2\left(\frac{m_2^* k_1}{m_1^* k_2} - \frac{m_1^* k_2}{m_2^* k_1}\right)^2 \sin^2(2k_1 L) \qquad (14)$$

The transmission and reflection coefficients are related by:

$$R + T = 1 \qquad (15)$$

The transmission coefficient T is given by the following form

$$T = \frac{\mu E_2\left[E_2 + 1 - (\mu-1)E_\parallel\right]}{\mu E_2\left[E_2 + 1 - (\mu-1)E_\parallel\right] + \left[\frac{\mu + (\mu-1)(E_2 - \mu E_\parallel)}{2}\right]^2 \sin^2\left[2k_{01}L\sqrt{1 + E_2 - (\mu-1)E_\parallel}\right]} \qquad (16)$$

where $\mu$ is the mass ratio $(\mu = m_2^*/m_1^*)$, the normalized initial electron energy in the z-direction of the structure is given by $E_i = \hbar^2 k_i^2 / 2m_i^* V_0$ and the normalized initial electron energy in the in-plane direction is given by $E_\parallel = \hbar^2 k_\parallel^2 / 2m_i^* V_0$. The coefficient T gets the larger value when the following equation is satisfied

$$L_n = n\pi \left(\frac{\hbar^2}{8m_1^*}\right)^{\frac{1}{2}} \left(V_0 + \frac{\hbar^2 k_2^2}{2m_2^*} - \left(\frac{m_2^*}{m_1^*} - 1\right)\frac{\hbar^2 k_\parallel^{(2)2}}{2m_2^*}\right)^{-\frac{1}{2}} \quad (17)$$

Fig 2 illustrates the variation of the transmission coefficient when the initial kinetic energy is $E_2$ with the in-plane energy $E_\parallel^{(2)} = E_2$ for an AlN/GaN quantum well. For the GaN/AlN heterostructure the effective mass ratio is $\mu = m_2^*/m_1^* = 3.2$. The well width is fixed in turn near the resonance points $48.65 \overset{\circ}{A}$ and $60.81 \overset{\circ}{A}$, so the chosen widths are $50 \overset{\circ}{A}$ and $60 \overset{\circ}{A}$. As it is clear, the transmission coefficient is strongly dependent on the free kinetic energy $E_\parallel^{(2)}$. Changing the ratio of the effective masses $\mu$ which influences the free kinetic energy, leads to the conclusion that the coefficient T becomes explicitly dependent on the free kinetic energy.

The final electron wavefunctions (with $E < V_0$) in the heterostructure which satisfy the Schrödinger equation can be given by:

$$\Psi_n^{(S)}(z) = B_S \begin{cases} \cos(k_1' z) & |z| \leq L \\ \cos(k_1' L) e^{-k_2'(|z|-L)} & |z| \geq L \end{cases} \quad (18)$$

$$\Psi_n^{(A)}(z) = B_A \begin{cases} \sin(k_1' z) & |z| \leq L \\ \sin(k_1' L) e^{-k_2'(|z|-L)} Sgn(z) & |z| \geq L \end{cases} \quad (19)$$

where the index S (A) denotes the symmetric (antisymmetric) electron wavefunctions. $k_1'$, $k_2'$ are the final electron wavevectors for region 1 and region 2 respectively. *Sgn(z)* is the sign of *z*. The coefficients $B_s$ and $B_a$ are calculated by using normalization of symmetric and antisymmetric wavefunctions respectively and by using the boundary conditions (continuity of $\psi(z)$ and $\frac{1}{m^*}\frac{d\psi}{dz}$ at the interfaces), the final electron states can be calculated.

The laser performance, among others, depends on carrier capture mechanism[11-12], the thermal broadening of the carrier populations and LO-phonon bottleneck effect [22]. In the last part of the current investigation, we theoretically descript the capture mechanism. Here, we consider an electron flux which is captured within the QW by the emission of hybrid phonons. Using the Fermi Golden Rule, we introduce a quantity which has dimensions of *"velocity"* and is called surface capture velocity [15d, 16]. Using the same manner as in ref. [16], the capture velocity gets the form

$$U_{Capt.} = \frac{4\pi m_1^*}{\hbar^3} \sum_{E_f} \sum_{j=1}^{N} \int_{\substack{|F(q_\parallel)|<1 \\ q_\parallel \geq 0}} dq_\parallel \frac{|ME'|^2}{k_\parallel^i (1-F^2(q_\parallel))^{\frac{1}{2}}} \quad (20)$$

where the function $F(q_\parallel)$ and the matrix elements $|ME|$ ($|ME'|$: dimensionless matrix elements) are respectively given by the following equations

$$F(q_\parallel) = \frac{1}{k_\parallel^i q_\parallel}\left[ q_\parallel^2 + \frac{2m_1^*}{\hbar^2}\left(\hbar\omega_j(\mathbf{q}_\parallel) - \Delta E^{(0)}\right)\right] \quad (21)$$

$$|ME|^2 = |\langle \Psi_i | e\Phi | \Psi_f \rangle|^2 = |A|^2 |ME'|^2 \quad (22)$$

where $\Phi$ is the scalar potential related to phonon modes [19]. The indices *i* and *f* correspond to the initial and final electron states respectively.

The capture velocity for an electron flux with energy in the continuum energy spectrum of the heterostructure from which electrons make transitions into the well by the emission of optical phonons, is the suitable parameter to measure the ability of the quantum wells to lase. The next task of our investigation is the numerical results of the capture velocities ($U_{Capt} = \Gamma_{Capt} L_0$, where $\Gamma_{Capt}$ are the capture rates and $L_0$ is the electron mean free path [16]) and the comparison with the reported experimental results.

## 3. Results

The capture velocity, as given by equation (20), depends on the scalar potential $\Phi$ representing the macroscopic phonon model that is used. Here the polar optical phonons are described by the hybrid modes. The capture velocity mediated by the emission HB phonons is illustrated in Fig. 3 and we compare these results to the DC model prediction [16]. The results clearly show resonances at specific QW widths. When the electron energy difference is equal to a phonon energy, a resonance appears and is called phonon resonance. Furthermore, another resonance kind appears when a new bound state is generated in the quantum well and is called electron resonance. At these peaks another condition takes place: at the electron resonance peaks the transmission coefficient in equation (16) has a large value. The phonon resonance peaks appear earlier than the electron resonance peaks due to the fact that the energy conservation is fulfilled before the generation of a new bound state. The height of the electron resonance increases linearly with increasing well width due to an increase in the number of electron states bound in the well by increasing the well width. Furthermore, another peak appears close to the main phonon resonance peak because of the number of the hybrid phonon modes which are involved in the scattering mechanism. Here, it should be noted that the in-plane energy ($E_\parallel$) is of crucial importance for the transmission coefficient and the well width where the electron resonances appears. So the choice of the in-plane energy will produce a different picture of the resonance magnitudes and the position of the resonance peaks.

The DC and the hybrid model predict almost the same results for the phonon resonances and the same electron resonances (see Fig. 3) due to the fact that the electron resonances do not depend on the model which describes the phonons in the heterostructures. It is worth noting that the DC model and the bulk phonon approximation predict close results for very small and large well widths [18]. As a result, all the above mentioned phonon models agree for the extreme cases of small or large widths. For the sake of simplicity, in the extreme cases, the simplest phonon model (phonon approximation) could bring reliable results for the electron capture processes.

Fig. 4 presents the average capture velocity $\langle U_{Capt.} \rangle$ as a function of the QW width for electron temperature T=300$^0$K and T=100$^0$K. The velocity $\langle U_{Capt.} \rangle$ was obtained by averaging over the Maxwell-Boltzmann distribution function of the electrons in a continuous spectrum with electron temperature T. We have chosen specific electron temperatures due to the existed available experimental results [11-12]. For the above conditions, the capture times are a few hundred of femtoseconds which are close to experimental values (200-600 fs).

## 4. Conclusions

In this article, we have studied the capture mechanism for a specific type of heterostructures and we have introduced the surface capture velocity for an electron flux which is captured into the QW. The capture velocity is a suitable parameter to compare and assess different quantum well systems that capture electrons from the same initial electron flux. The quantum process is assisted by the emission of LO phonons as described by the HB model. The results show that the capture velocity exhibits sharp peaks at regular intervals with increasing quantum well width corresponding to electron resonances in the same manner as evaluated in ref. [15] where the initial electron state is well defined. The smaller sharp peaks emerge whenever the electron energy becomes equal to a phonon energy.
Furthermore, the comparison of the DC and HB phonon models (Fig. 3) shows that the two models predict almost the phonon resonances and the same electron resonances number. The sharp peaks in the capture velocities show that lasing emission should be more pronounced at specific well widths. Among other calculated parameters [22], the evaluation of these well widths is one of the most important calculations for choosing suitable laser heterostructures. It is worth noting that nitride QWs have more resonance peaks than less polar materials e.g. GaAs[15 (d)]. Lastly, the comparison of our results to the recent experimental capture times shows that the theoretical and experimental results are in good agreement.

5. **Acknowledgements**

The authors would like to thank Prof. M. Babiker, Prof. Brian Ridley for useful discussions and Commander of Hellenic Navy D. Filinis for his support. The author V.N.S. would like to acknowledge the financial support given by the Training Mobility of Researchers (TMR) and the University of Iowa under the grand No. 52570034.

**Figure captions**

**Figure 1.** The profile illustration of a typical symmetric quantum well of width 2L made from materials 1, 2 and 3.

**Figure 2.** The transmission coefficient as a function of the initial kinetic energy $E_2$ (with $E_{\parallel}^{(2)} = E_2$ and $E_2$ is normalized to $V_o$.) for an AlN/GaN quantum well width $50 \overset{0}{A}$ (solid lines) and well width $60 \overset{0}{A}$ (dotted lines) for different effective mass ratios.

**Figure 3.** The capture velocity versus the well width with $E_{\parallel}^{(2)} = 0$ and total energy $E = \dfrac{\hbar\omega_{L2}}{2}$ for an AlN/GaN quantum well for the DC and HB phonons.

**Figure 4.** The average capture velocity as a function of well width for a quantum well made with AlN/GaN.

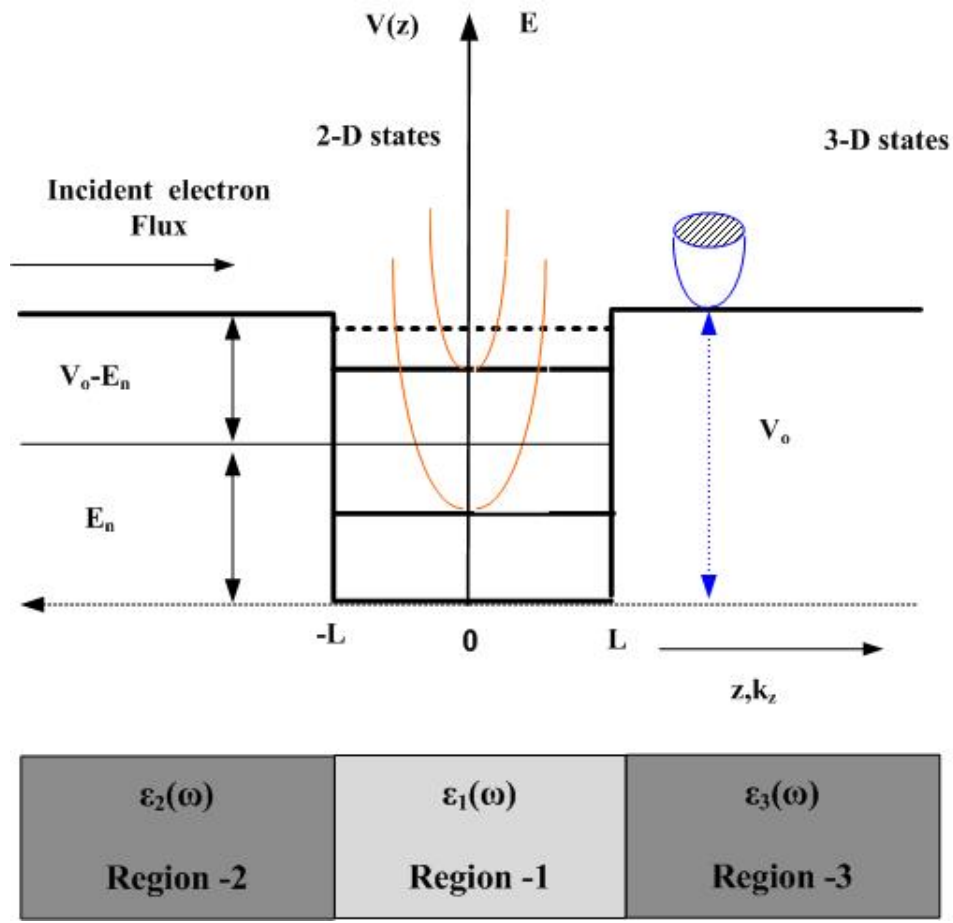

**Figure 1.**  V. N. Stavrou & G.P. Veropoulos

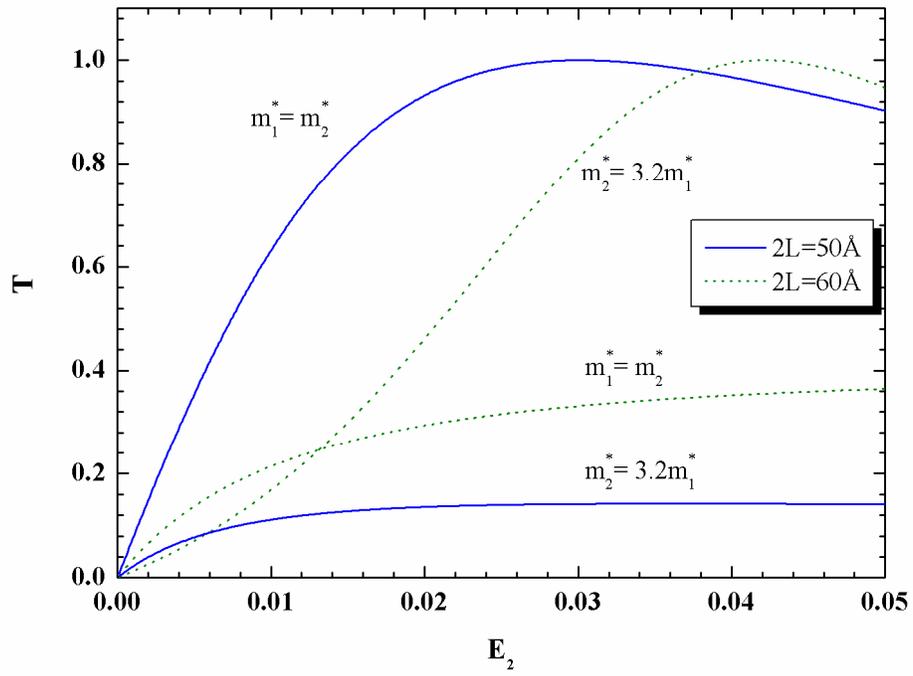

**Figure 2.**     V. N. Stavrou & G.P. Veropoulos

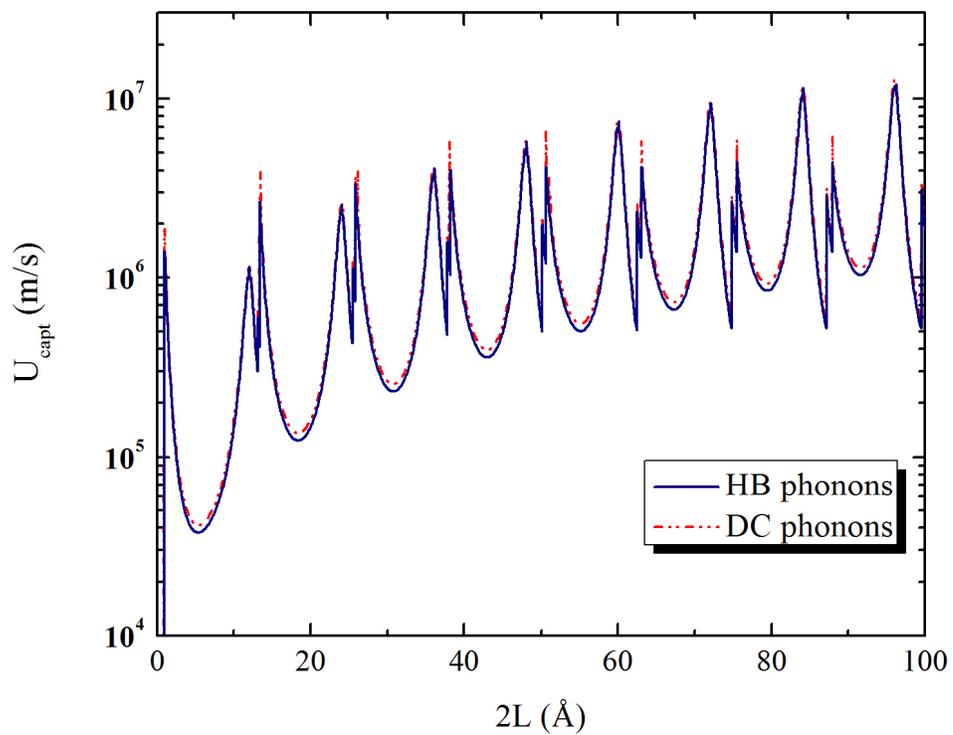

**Figure 3.**      V. N. Stavrou & G.P. Veropoulos

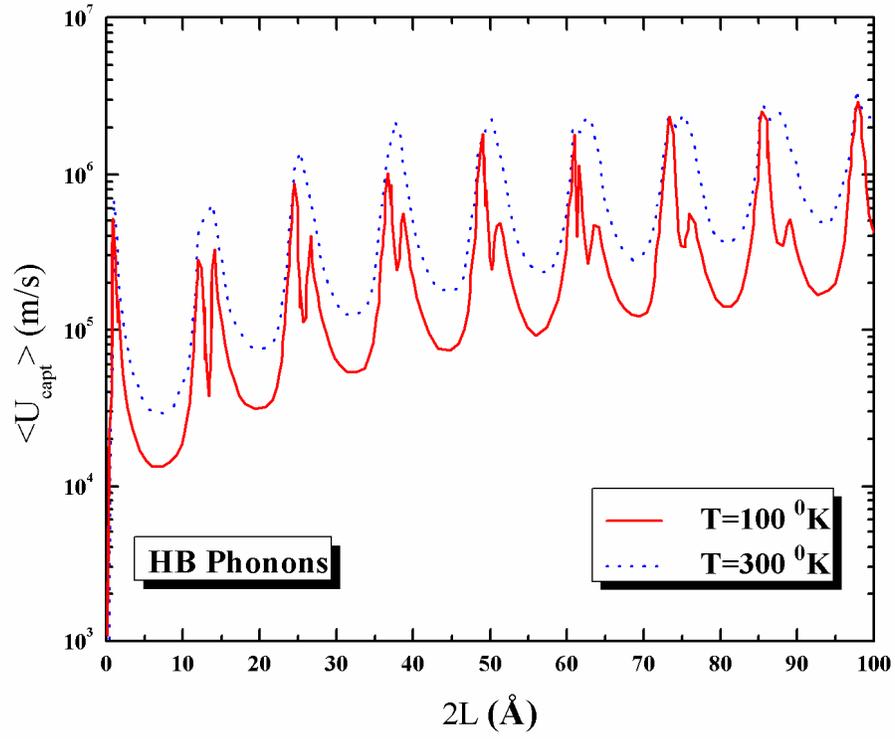

**Figure 4.**     V. N. Stavrou & G.P. Veropoulos